\documentclass[conference]{IEEEtran}
\IEEEoverridecommandlockouts

\usepackage{cite}
\usepackage{amsmath,amssymb,amsfonts}
\usepackage{algorithmic}
\usepackage{graphicx}
\usepackage{textcomp}
\usepackage{caption}
\usepackage{subcaption}
\usepackage{xcolor}
\usepackage{soul}

\def\BibTeX{{\rm B\kern-.05em{\sc i\kern-.025em b}\kern-.08em
    T\kern-.1667em\lower.7ex\hbox{E}\kern-.125emX}}
\begin{document}

\title{Transient Performance Modelling of 5G Slicing with Mixed Numerologies for Smart Grid Traffic\\
}

\author{
\IEEEauthorblockN{
H. V. Kalpanie Mendis\IEEEauthorrefmark{1},
 Poul E. Heegaard\IEEEauthorrefmark{1},
 Vicente Casares-Giner\IEEEauthorrefmark{2},
 Frank Y. Li,\IEEEauthorrefmark{3}
 and Katina Kralevska\IEEEauthorrefmark{1}
} 
\IEEEauthorblockA{\IEEEauthorrefmark{1}Department of Information Security and Communication Technology,\\
NTNU - Norwegian University of Science and Technology, N-7491 Trondheim, Norway \\ 
\IEEEauthorrefmark{2}Department of Communications, Universitat Politècnica de València, València 46022, Spain\\
\IEEEauthorrefmark{3}Department of Information and Communication Technology, 
University of Agder (UiA), N-4898 Grimstad, Norway\\ 
Email: \{kalpanie.mendis; poul.heegaard; katinak\}@ntnu.no; vcasares@dcom.upv.es; frank.li@uia.no } } 

\maketitle

\begin{abstract}
Network slicing enabled by fifth generation (5G) systems has the potential to satisfy diversified service requirements from different vertical industries. As a typical vertical industry, smart distribution grid poses new challenges to communication networks. This paper investigates the behavior of network slicing for smart grid applications in 5G radio access networks with heterogeneous traffic. To facilitate network slicing in such a network, we employ different 5G radio access numerologies for two traffic classes which have distinct radio resource and quality of service requirements. Three multi-dimensional Markov models are developed to assess the transient performance of network slicing for resource allocation with and without traffic priority. Through analysis and simulations, we investigate the effects of smart grid protection and control traffic on other types of parallel traffic sessions as well as on radio resource utilization.  
\end{abstract}


\section{Introduction}\label{sec:intro}
Network slicing is a novel paradigm in fifth generation (5G) systems which facilitates multiple virtual end-to-end (E2E) networks by reserving resources based upon one physical network infrastructure. A network slice is a logical network with specific network capabilities and network characteristics, tailored to offer customized communication services. For instance, different slices could be defined to satisfy the diverse service requirements imposed by the three 5G use cases, i.e., enhanced mobile broadband (eMBB), ultra-reliable low latency communications (URLLC), and massive machine-type communications (mMTC).   

An E2E network slice is composed of an access network (wired, or wireless radio access network (RAN)), and a core network(s) (sub-)slices between end users or devices. Recently, there is a surge of interests in RAN-based network slicing from both academia and standardization bodies~\cite{RAN_COMMAG19}~\cite{RAN_COMMAG17}. For instance, the 3rd generation partnership project (3GPP) is currently investigating RAN support of network slicing \cite{TR38832}. In this paper, we are interested in radio resource allocation in a RAN, which  constitutes a part of an E2E network slice, referred to as a RAN slice. The numerology design in 5G new radio (NR) facilitates the flexibility of customized RAN slice design according to the heterogeneous requirements of vertical applications.

Considering the dynamic nature of network traffic, adaptive resource sharing can make network resource utilization more efficient. An approach for radio resource virtualization and RAN resource allocation was proposed in~\cite{infocomws18}. Targeting at two RAN slices for eMBB and URLLC services,~\cite{RANslicing_gc19} proposed a two-level medium access control (MAC) scheduling algorithm for radio resource sharing and dynamic adjustment. However, resource sharing also brings challenges for slice isolation. Accordingly, a proof-of-concept prototype for RAN run-time slicing which has the capability of isolating, sharing, and customizing resources for three use cases was demonstrated in~\cite{Access0618}. Based on a flexible numerology structure in 5G NR, dynamic resource allocation was formulated as an optimization problem for quality of service (QoS) provisioning given the co-existence of eMBB and URLLC services in \cite{tang2019dynamic}. Moreover, the authors in~\cite{Gonzmixnum} presented a quality of service (QoS)-aware resource allocation scheme in multi-carrier systems supported by the flexibility obtained through mixed numerologies. 

Considering both guaranteed and non-guaranteed bit rate services, a continuous time Markov chain model for RAN slicing was presented in~\cite{Access0320}, targeting at characterizing diverse radio resource management strategies in multi-tenant and multi-service 5G scenarios. However, no Markov model which analyzes network slicing and resource allocation at the frame or subframe level in 5G NR exists so far, especially when the transient behavior is of interest.

As network slicing offers context-related and customized services, its applicability to many vertical industries is promising. Power distribution grid is a typical example of a vertical industry as it is regarded to be a critical infrastructure in our modern society. In smart grid applications, power outages must be handled in milliseconds for power operator and consumers. Hence, mission-critical operations such as protection and control in smart grid cannot be accomplished without URLLC.

\begin{table*}[htbp] 
  \centering \small
  \caption{5G New Radio numerologies \cite{TR38101} \vspace{-2mm}}
  \label{tab:num}
  \begin{tabular}{ccccccc}
    \hline
    Numerology, $\mathcal{\beta}$ & SCS 
    [kHz] & PRB size 
    [kHz]   & \#slots/subframe 
    & Cyclic prefix (CP)  & Symbol duration [$\mu$s] & CP duration [$\mu$s]\\
    \hline
    0  & 15 & $180 $  & 1 & Normal  & $71.43$ & $4.69$\\
    1  & 30 & $360$  & 2 & Normal  & $35.71$ & $2.34$\\
    2 & 60 & $720 $  & 4 & Normal, extended  & $17.86$ & $1.17$\\
    3 & 120 & $1440$   & 8 & Normal & $8.92$ & $0.57$\\
    4 & 240  & $2880$  & 16 & Normal & $4.46$ & $0.29$\\
    \hline 
  \end{tabular} 
\end{table*}

In this paper, we apply 5G RAN slicing to smart grid applications with power distribution protection and evaluate its performance considering two types of traffic. More specifically, the offered traffic to a 5G RAN is categorized into two traffic classes based on their radio resource, priority, latency, and reliability requirements. For the mission-critical protection message type, the 5G URLLC traffic class is needed~\cite{mendis20195g}, and is treated in one RAN slice instance, whereas the background traffic type includes both power grid traffic with less restrictive requirements, and traffic from other non-critical services is considered as background traffic in another RAN slice. 
The radio link between end devices and their associated next generation NodeB (gNb) is considered to have limited capacity for each network slice and its resource utilization is assessed.

In a nutshell, the main contributions of this paper include 
(i) the proposed criteria for selecting the most preferable numerology for a set of services and under given network conditions; and 
(ii) performance assessment of 5G RAN slicing with hybrid numerologies to support heterogeneous traffic classes, \emph{focusing on the transient behavior of sliced networks}.  
To do so, three multi-dimensional Markov models are developed and applied to smart grid operations (which receive high priority due to their stringent low latency and high reliability requirements). 

The remainder of this paper is organized as follows. Sec.~\ref{sec:num} summarizes 5G NR and numerology schemes, and introduces various message types that are defined for grid protection and control. Then the network scenario is presented in Sec.~\ref{sec:sys}, and the proposed multi-dimensional Markov models are presented in Sec.~\ref{sec:models}. Afterwards, qualitative and quantitative results are presented in Sec.~\ref{sec:assess}. Finally, Sec.~\ref{sec:con} concludes the paper.


\section{Preliminaries}\label{sec:num}
5G NR defines a flexible time-frequency lattice,  enabling a multi-numerology structure. Thanks to this flexibility, NR is capable of providing services across diverse usage scenarios.

\subsection{5G New Radio}
5G frequency range 1 (FR-1) includes sub-6GHz frequency bands and is envisaged to carry much of the traditional cellular communication traffic in low traffic density areas. At higher frequency bands from 24.25 GHz to 52.6 GHz, frequency range 2 (FR-2) aims at providing ultra-high data rate capability within short ranges in high density traffic areas. As 5G NR can be operated in both wider and narrower channel bandwidth, the orthogonal frequency division multiplexing (OFDM) sub-carrier spacing (SCS) has to scale accordingly so that the complexity of fast Fourier transformation (FFT)  does not increase exponentially with wider bandwidth. 

Compared with long term evolution (LTE) and LTE-advanced (LTE-A) that support carrier bandwidth up to 20~MHz with a fixed 15 kHz OFDM SCS, 5G NR offers scalable OFDM numerology to support diverse spectrum bands and deployment modes. This is achieved by enabling multiple numerologies obtained by multiplying the basic LTE/LTE-A SCS with $2^\beta$ where $\beta$ is an integer between 0 and 4. Tab.~\ref{tab:num} presents the main features of each of the 5 numerology structures defined in 5G NR~\cite{TR38101}. 

The resource grid of 5G NR comprises of a set of physical resource blocks (PRBs). A PRB in NR is defined as the product of 12 consecutive SCSs in the frequency domain and one OFDM symbol in the time domain. For downlink traffic, one PRB is the smallest frequency-time unit that can be assigned to a device or user equipment (UE).  
The same as in LTE/LTE-A, one NR frame contains 10 sub-frames each lasting for 1 ms. With different number of slots within one sub-frame, flexible resource allocation with various granularity is enabled. However, the number of OFDM symbols within one slot does not change with numerology and it remains as~14. A contiguous set of PRBs selected for a given numerology is referred to as a bandwidth part (BWP). 

A network slice may contain a certain number of BWPs and PRBs. Depending on the amount of radio resources a service provider owns as well as the types of services and traffic volume, it may define a network slice for a specific type of service or a common slice for multiple types of services. Both static and dynamic slice configuration are possible. 

\subsection{Communication Requirements in Smart Distribution Grid}\label{subsec:use-case}

Various smart distribution grid operations impose diverse communication service requirements in terms of latency, bandwidth, reliability, security, etc. Therefore, smart grid applications may benefit from a scalable and flexible communication architecture which can be realized through 5G network slicing. A few power grid operation examples are analyzed and multiple service requirements are specified in~\cite{3gppTS22104}. 

Intelligent electronic device (IED) refers to any device that has processing and communication capabilities. These devices are expected to exchange information with a control panel and to make decisions based on the information they receive. 

Furthermore, international electrotechnical commission (IEC) 61850 is a standard that defines a set of communication protocols and abstracts data models for IEDs. The IEC 61850 standard defines two real-time, peer-to-peer communication methods: sampled values (SV) messaging and generic object oriented substation event (GOOSE) messaging. For measurement-type communications, SV messages are adopted and the data transmission frequency is pre-configured by the distribution grid operator. On the other hand, GOOSE is typically event driven and this type of messages will be generated when an event (e.g., alarm, failure, or any mission-critical event) occurs. In~\cite{mendis20195g}, different message types used for distribution grid protection and control are categorized according to two key performance requirements (availability and latency) and mapped onto the three major 5G use cases presented above. 

GOOSE messaging is operated based on a publisher-subscriber mechanism with its messages broadcast over a local network to all devices that are listening to or have subscribed to the network. For achieving higher reliability for information dissemination, a GOOSE message may be sent more than once. To reduce latency, no acknowledgement mechanism is introduced. Therefore, URLLC service is expected in order to ensure reliable operation as well as to meet the real-time requirement of GOOSE transmissions.

When there is no GOOSE event, a publishing IED sends heartbeat messages at a time interval of $t_{\tt max}=1000\;[\tt ms]$ to all subscribing IEDs. If a receiving IED does not receive a GOOSE message within $2\cdot t_{\tt max}$, it will regard that the sending IED as being offline~\cite{huang2017practical}. The publishing IED which notices a failure, for instance in a feeder, will change the payload of the GOOSE message accordingly and send it instantly. The IED will go to a mode referred to as the fast repetition mode and will send the above GOOSE message repeatedly every $5\; [\tt ms]$ until the network has been recovered.


\section{Network Scenario}\label{sec:sys}
\begin{figure}[t]
\centering 
\includegraphics[width=1\columnwidth]{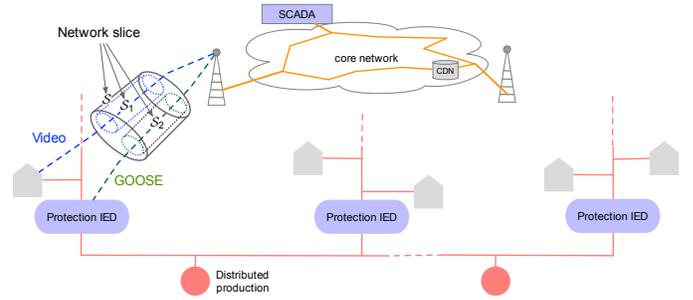}
\caption{\small{Illustration of the network scenario. SCADA stands for supervisory control and data acquisition.}} 
\label{fig:usecase}
\end{figure}

In this study, we focus on a common 5G NR infrastructure where the radio resources belonging to the same RAN (considering a  service provider with a common network infrastructure) are shared by two types of services with multiple tenants. Two network slices are available and are managed by (a single) management and orchestration (MANO) entity. The envisaged scenario is illustrated in Fig.~\ref{fig:usecase} where two network slices share the same \emph{physical} resources but these two types of services are separated by two \emph{logical} network slices. The considered two traffic types are GOOSE messages (for smart distribution grid protection and control traffic) served by slice ${\tt S}_1$ and interactive video sessions (either adaptive or non-adaptive) served by slice ${\tt S}_2$, respectively. Three network configurations (NCs) are investigated in this study, as presented below. 

\begin{itemize}

\item NC1: Both traffic types are mapped onto a shared network resource pool (shown in Fig. \ref{fig:usecase}) and no priority is given to the GOOSE traffic. 

\item NC2: Within the shared resource pool, higher priority is given to GOOSE messages and coexisting video sessions are non-adaptive. By non-adaptive, it is meant that a video session always runs at a full rate and it will be dropped or rejected if no  sufficient resource is available.     

\item NC3: The same as in NC2 but coexisting video sessions are adaptive. By adaptive, it is meant that a video session may be operated at a downgraded rate (at a half rate in this work) with less resources instead of being dropped as in NC2. It may be dropped if the resource is still insufficient for downgraded flows.    
\end{itemize}

In a multi-tenant environment with a shared RAN, the physical or virtual infrastructure resources may be dedicated to a network slice or shared with other network slices. In order to guarantee performance isolation and to avoid service interruptions, resources maybe pre-reserved for a specific network slice. However, it is worth mentioning that in our scenario no resources are reserved to the GOOSE traffic so that the idle resources in ${\tt S}_1$ can be utilized by other types of traffic (video, background, etc.) if no GOOSE event occurs.

\section{Markov Models}\label{sec:models}
In this paper, we apply Markov loss models~\cite{Iversen:2005} to study and quantify the cross-dependence between network slices hosting GOOSE and video sessions. We compare configurations with and without priority between the slices (priority of GOOSE over video). In what follows, we present first the basic loss model, and then develop three models which are dedicated to the three network configurations explained above. Afterwards, a number of performance metrics are defined.

\subsection{Generic System Model and $n$-Dimensional Loss Model}\label{subsec:system-model}

To develop Markov models for the system with three NCs presented in Sec.~\ref{sec:sys}, we have defined the system state as $\omega = \{ \omega_1, \cdots, \omega_n \}$, where $\omega_i \in [0;s_i]$ is the number of active sessions of service type $i$, $s_i$ the maximum number of session of type $i$, and $n$ the number of traffic types.  
The $n$-dimensional state space is denoted by $\Omega$.  
The number of active sessions at time $t$ is $\mathbf{M}(t) = \{ M_1(t), \cdots, M_n(t)\}$, $\mathbf{M}(t)\in \Omega$. 
These notations will be used in Subsec.~\ref{subsec:perf} when we define performance metrics.
$\lambda_i(\omega)$ is the arrival rate,  $1/\mu_i(\omega)$ is the expected session duration, 
$\delta_i$ is the number of resource blocks per session, and lastly $C$ denotes the total number of resource blocks (i.e., the system capacity).

All times in the system are assumed to be exponentially distributed, i.e., the events are generated by Poisson processes. Hence, the system models in this paper are Markovian. Although the effect of this assumption may be further investigated, we believe that the results obtained based on our models provide valuable insight on network behavior, \emph{especially during the transient period when a GOOSE burst is the injected}. The GOOSE bursts are modelled as a train of GOOSE packets with their interarrival times \emph{much shorter} than the time between video sessions, and hence, the consequence of assuming exponential instead of any other general distribution is negligible.

Based on the generic system model, we develop a basic $n$-dimensional loss model, where each dimension is a traffic type. A type is either defined by the slice ($k\le n$) it belongs to, or a traffic type (e.g., full rate or downgraded rate video).
An arrival of a new session of type $i$ will be rejected if no sufficient capacity (in terms of resource blocks) is available, i.e., $\sum_{j=1}^n \delta_{j} (\omega_j+1_i) > C$, where $1_i=1$ when $j=i$, and $0$ otherwise.

Note that the model presented in this subsection is generic and it can be applied to multiple types of services. In what follows, we develop three models where $n=2$ for NC1 and NC2 and $n=3$ for NC3 respectively, dedicated to the three network configurations presented in Sec. \ref{sec:sys}. 

\begin{figure}[t]
\centering
\includegraphics[width=1\columnwidth]{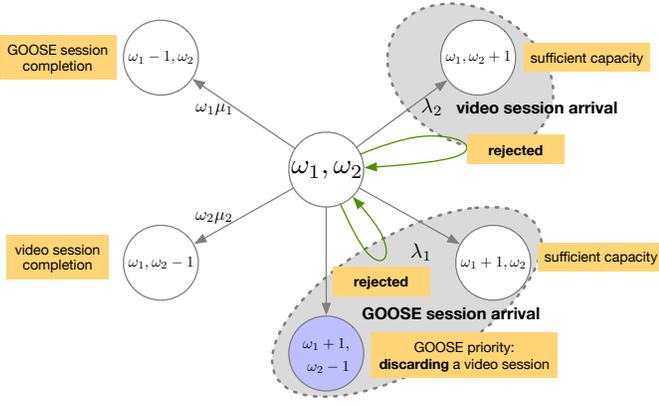} \vspace{-7mm}
\caption{\small{State $(\omega_1,\omega_2)$ of the model with non-adaptive video sessions.}} \vspace{-3mm} 
\label{fig:MMnonadapt}
\end{figure}

\subsection{Model for NC1: Non-Priority, Non-Adaptive Video Traffic}\label{subsec:nonpri}
A system with two traffic classes (GOOSE and non-adaptive video) and no priority between these classes (e.g., they are mapped onto the same network resource pool) can be modelled by the generic Markov model introduced above.  

The general state $(\omega_1,\omega_2)$ of this Markov model is illustrated in Fig.~\ref{fig:MMnonadapt} (with the transition to the state ``GOOSE priority'' disabled).  The yellow labels explain the events and the conditions that cause the transition.
$\lambda_1$ is the GOOSE session arrival rate and $\lambda_2$ is the video session arrival rate, while $\mu_1$ and $\mu_2$ are the session departure rates of GOOSE and video traffic, respectively. Note that if there is not sufficient capacity~$C$ upon the arrival of either class, the session is rejected, e.g., if $(\omega_1+1)\delta_1 + \omega_2 \delta_2 > C \ge \omega_1\delta1+\omega_2\delta2$, a GOOSE session arrival will be rejected. This is due to the fact that in NC1 no priority is given to the GOOSE traffic.

\subsection{Model for NC2: Priority, Non-Adaptive Video Traffic}\label{subsec:prinonadapt}
To model a system with two traffic classes (GOOSE and non-adaptive video) but now with priority on the GOOSE sessions (e.g., each class has its own network slice), the Markov model in Fig.~\ref{fig:MMnonadapt} is slightly updated. If  a GOOSE session arrives in a state with insufficient capacity to accommodate both GOOSE and video session, then the system will discard an ongoing video session and accept the GOOSE session. If all PRBs are occupied by GOOSE sessions, new arrivals (both video and GOOSE) will be rejected.  This is illustrated by the transition to the state ``GOOSE priority'' (the state marked in purple) in the same figure.

\subsection{Model for NC3: Priority, Adaptive Video Traffic}\label{subsec:priadapt}

To avoid discarding too many ongoing video sessions, an alternative is to downgrade their service rates. This means that less resource blocks are needed per session (i.e., $\delta_i$ is reduced) after downgrading.  The consequence is either a lower quality (by changing encoding scheme for instance to an embedded scheme with lower resolution) or increased download times.  

\begin{figure}[t]
\centering \vspace{-6mm}
\includegraphics[width=1\columnwidth]{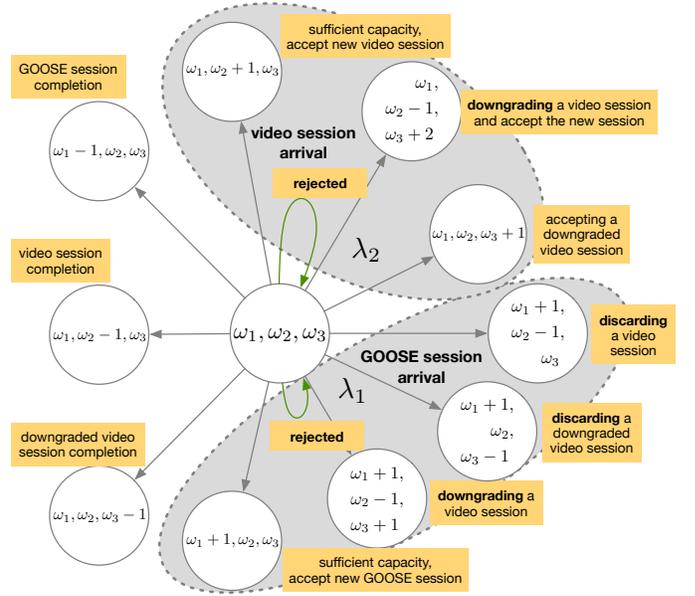} \vspace{-10mm}
\caption{\small{State $(\omega_1, \omega_2, \omega_3)$ of the Markov model of the high priority GOOSE sessions and adaptive video sessions.}} \vspace{-3mm}
\label{fig:MMadapt}
\end{figure}

To model the behavior of NC3 where both full rate and downgraded rate video sessions coexist, we extend the model presented above by introducing one more dimension to form a 3-dimensional model $(\omega_1,\omega_2,\omega_3)$, where $\omega_3$ is the number of downgraded video sessions and  $\delta_3 < \delta_2$. With this update, the arrival of a GOOSE session will then be accepted despite insufficient capacity, since we can downgrade a video session instead of discarding it.  If the system capacity is still insufficient after downgrading all ongoing video sessions, a GOOSE session could still be rejected. Similarly, upon the arrival of a video session to a state with insufficient capacity, the system first tries to downgrade it before it is rejected. A detailed transition diagram for this model is illustrated in Fig.~\ref{fig:MMadapt}. See the yellow labels in the figure for further explanations.

\subsection{Performance Metrics}\label{subsec:perf}
For performance comparison among the different configurations, a set of performance metrics are defined. Recall that
$M_i(t)$ is the number of sessions of type $i$ at time $t$ and $c_i(t) = M_i(t) \delta_i$ is the number of resource blocks occupied by type $i$ at time~$t$. This gives the total number of resource blocks at time $t$ as $C(t) = \sum_{i=1}^n c_i(t)$. Accordingly, \emph{resource block utilization}, denoted by $\rho(t)$, is defined as the number of physical resource blocks normalized by the total capacity, i.e., $\rho(t) = C(t)/C$ ($\rho$ is the utilization when $t \to \infty$).

A GOOSE burst duration ($\mathcal{T}$) is the time period from the instant when the first GOOSE session enters the system to the instant when all GOOSE sessions have left the system. To measure the impact of GOOSE burst arrivals on video traffic, we define three other metrics, as the \emph{ratios} of the number of rejected, downgraded, and discarded video sessions and the number of arrivals over the observation period ($n_{\tt ga}$). 
These ratios are denoted by $r_{\tt rj}$, $r_{\tt dw}$, and $r_{\tt dc}$, respectively. Furthermore, the ratio of a rejected video session (without GOOSE) is $r_{\tt v}$. 

To study  the transient effect of GOOSE sessions, we further define a {\em GOOSE burst period}, denoted by $\mathcal{T}_{\tt g}$, as the time duration from injecting $s_{\tt g}$ sessions at time instant $T_{\tt I}$ until they are completed, $\mathcal{T}_{\tt g} = \max(\{t | M_{\tt g}(t)>0\}) - T_{\tt I}$, where $M_g(t)$ is the number of GOOSE sessions at time $t$.

\begin{figure}[t]
\centering
\includegraphics[angle=0,width=1\columnwidth]{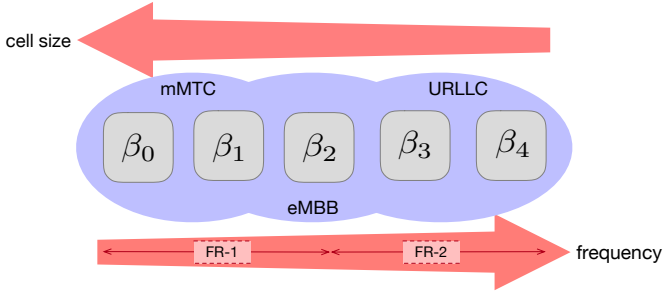}
\vspace{-3mm}
\caption{\small{Criteria for  numerology selections.}} \vspace{-3mm}
\label{fig:num_select}
\end{figure}

\subsection{Simulation of Markov Models}\label{subsec:MMsim}
From the Markov models developed above, the performance metrics introduced in Subsec.~\ref{subsec:perf} can be obtained. For steady state analysis, solutions such as Kaufman-Roberts~\cite{Kaufman:1981}~\cite{Roberts:1981}  
can be applied (in our case for the non-priority case, NC1).  

For transient solutions which is the focus in the paper, we apply simulation of the Markov model~\cite{Jensen:1953}.  Performing Markov simulations based on the models we developed above (shown in Fig.~\ref{fig:MMnonadapt} and Fig.~\ref{fig:MMadapt} respectively) is a viable alternative for performance evaluation of transient network behavior. Obtaining closed-form analytical expressions for transient behavior is rather complicated or intangible, and it is thus beyond the scope of this paper. See Subsec. \ref{sec:quant} for more details on simulation of Markov chains.

\section{Qualitative and Quantitative Assessments}\label{sec:assess}
This section discusses selection of numerology qualitatively and presents the network performance quantitatively.

\begin{figure*}[tb]
\begin{subfigure}{\columnwidth}
  \centering
  \includegraphics[width=1\columnwidth]{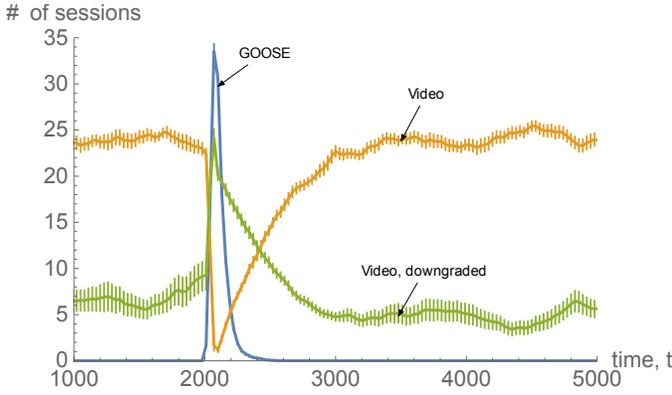}  \vspace{-1mm}
  \caption{$\lambda_{2}=1/20$ [s$^{-1}$], $\rho=0.76$, $r_{\tt v}=0.11$} \vspace{-1mm}
  \label{fig:sub-first}
\end{subfigure}
\begin{subfigure}{\columnwidth}
  \centering
  \includegraphics[width=1\columnwidth]{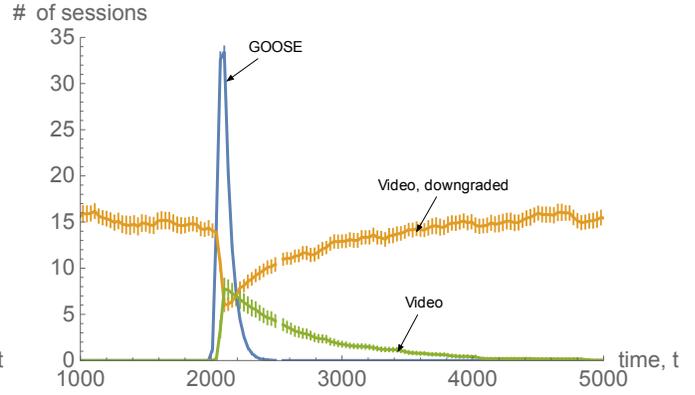}  \vspace{-1mm}
  \caption{$\lambda_{2}=1/40$ [s$^{-1}$], $\rho=0.49$, $r_{\tt v}=0.003$} \vspace{-1mm}
  \label{fig:sub-second}
\end{subfigure}
\caption{\small{Number of sessions over time, $M_i(t)$, resource block utilization $\rho$, and rejected video sessions $r_v$.}}
\vspace{-3mm}
\label{fig:sessions}
\end{figure*}

\subsection{Qualitative Considerations of Numerology Selection}
One objective of this paper is to investigate the tradeoffs to be made in mapping traffic classes to 5G numerologies with respect to resource utilization with network slicing.

\subsubsection{Mapping traffic classes to numerologies}
In general, the selection of numerology will depend upon various factors including type of deployment (i.e., dense urban, rural, urban macro etc.), carrier frequency, service requirements (latency, reliability, and throughput), hardware impairments (oscillator phase noise), mobility and implementation complexity. Note further that not all five SCS options are applicable to all frequency bands~\cite{TR38101}~\cite{peisa20175g}.

High numerologies can be advantageous for latency-critical services such as URLLC since wider SCS implies a shorter transmission time interval (TTI). Deployments with smaller cell sizes using higher frequency bands are affected from high levels of phase noise. The phase noise can be mitigated using a wider SCS. Moreover, for scenarios with higher Doppler spread, which is affected by mobility, multipath propagation and angular speed, a larger SCS needs to be adopted. Lower numerologies, on the other hand,  can be utilized for narrowband devices implemented for mMTC use cases. Large cells have a large time dispersion (long delay spread), thus benefiting from narrower SCS. A summarized illustration for selecting a suitable numerology is given in Fig.~\ref{fig:num_select}.

However, our observation is that the number RBs allocated for each session is the same irrespective the chosen numerology, thus does not change the Markov model or the simulation results presented later.

\subsubsection{Guard band and guard period overheads}
Isolation is one of the key requirements for network slicing deployment. A network slice instance (NSI) which stands for an activated slice may be fully or partly, logically and/or physically, isolated from other NSIs~\cite{TR28801}. In order to ensure the isolation among RAN slices, the resources allocated to one NSI must not overlap with that of other NSIs. 

Although 5G NR improves spectral efficiency, it also introduces non-orthogonality to the system, causing interference among different numerologies, which is known as inter-numerology interference (INI). Thus, the flexibility of multi-numerology structure comes at a cost of computational complexity and signal overhead. To minimize INI, solutions with adaptive guards in both time and frequency domains may be applied~\cite{demir2020inter}. 

3GPP~\cite{TR38101} specifies the minimum guard band, $\mathcal{G}$, for each UE channel bandwidth and SCS, as $\mathcal{G} = (C - \delta \cdot {\tt SCS} \cdot 12)/2$. For any given bandwidth, a gNB shall ensure that the above minimum guard band is met when allocating the number of PRBs in a cell. For our simulations presented in Subsec.~\ref{sec:quant}, we have considered  channel bandwidth of 25~[MHz], which requires 1.33~[MHz] guard bands on each side of the carrier.

Furthermore, when multiple numerologies are multiplexed inside the same subframe, a guard band or guard period has to be configured to mitigate INI. In 5G NR, two frame structures exist for frequency division duplex (FDD) and time division duplex (TDD) respectively. For FDD with full-duplex, no explicit guard period is needed. For FDD with half-duplex, it requires a guard period of one OFDM symbol. For TDD, a guard period may vary from one to ten OFDM symbols depending on downlink pilot time slot (DwPTS) and uplink pilot time slot (UpPTS) configurations. In this study, we do not consider signalling overhead or initial access as we concentrate on resource  allocation after the  handshake  phase. Since the required guard period between two numerologies is negligible in terms of the radio resource occupancy, we have ignored this overhead in our simulations.

\subsection{Quantitative Assessment of Cross-Slice Effects}\label{sec:quant}

Extensive simulations are performed to assess the cross-slice effects based on the three configurations presented above. The duration of the observation period is configured as 6 seconds. In the initial phase, only video sessions are carried out. Then GOOSE sessions are injected at $T_{\tt I} = 2000\;[\tt ms]$. The observation period terminates either when the observation time reaches $T = 6000\;[\tt ms]$ or when there are $s_1$ GOOSE sessions. The simulation is replicated $30$ times with the same initial and terminating conditions. Then the sample mean values and the sample variances are obtained and shown in Figs. \ref{fig:sessions} and~\ref{fig:burst} respectively. The simulation configurations are listed in Tab.~\ref{tab:sim para}. 

\begin{table}[t]
\centering
\caption{Simulation parameter configurations} \vspace{-2mm}
  \label{tab:sim para}

\begin{tabular}{llp{26mm}l} 
 Type of traffic, $i$                     & 1: GOOSE & 2: Video \\ 
\hline
Arrival rate, $\lambda_i$ [s$^{-1}$]          & Injection rate: $1$    & $1/10,1/20,1/40$ \\ 
Service rate, $\mu_i$ [s$^{-1}$]            & $1/60$     & $1/600$        \\ 
Numerology, $\mathcal{\beta}$            & $1$     & $2$        \\ 
Resources/session, $\delta_i~$[kHz] & $360$     &  $720$ (full-rate) or $360$ (downgraded)       \\
Max. no. of sessions, $s_i$ & $62$     & $31$       \\
Priority & High $|$ none     & Low $|$ none       \\
\hline
\end{tabular} 
\end{table}

\subsubsection{Number of sessions}
Due to the page limit, we present only the results for NC3 in Fig.~\ref{fig:sessions}. The curves therein illustrate how the number of sessions $M_i(t)$ varies over time for video full rate, downgraded rate, and GOOSE, respectively. When the number of full rate video sessions reaches the maximum number of sessions allowed, i.e., when the total system capacity is occupied, the newly arriving video sessions will be downgraded. That is when the downgraded video sessions appear on the plot. When the GOOSE sessions are injected into the network at $T_{\tt I} = 2000\;[\tt ms]$, a significant amount of full rate video sessions have to be switched to downgraded video. Some video sessions may be even discarded due to insufficient capacity. After the GOOSE bursts have left the network, the number of video sessions increases to fill up the capacity available in the network, released by GOOSE traffic.

Fig.~\ref{fig:sub-first}) and Fig.~\ref{fig:sub-second}) reveal the performance when the video arrival rate is $\lambda_{2}= 1/20$ [s$^{-1}$] and \textbf{$\lambda_{2}=1/40$} [s$^{-1}$], respectively. Clearly, the expected number of full rate video sessions is dependent on the arrival intensity. At a low arrival rate, the number of full rate sessions will be higher, and vice versa. At \textbf{$\lambda_{2}=1/40$} [s$^{-1}$], we observe that no video sessions need to be downgraded before a GOOSE burst occurs. When GOOSE sessions are injected, a number of video sessions need to switch to a downgraded video rate in order to accommodate GOOSE sessions. With a higher $\lambda_{2}$, the resource utilization, $\rho$, will be higher, at a cost of a higher number of rejected video sessions $r_{\tt rj}$. 

\subsubsection{Duration of the GOOSE bursts}

\begin{figure}[tb]
    \centering
    \includegraphics[width=1\columnwidth]{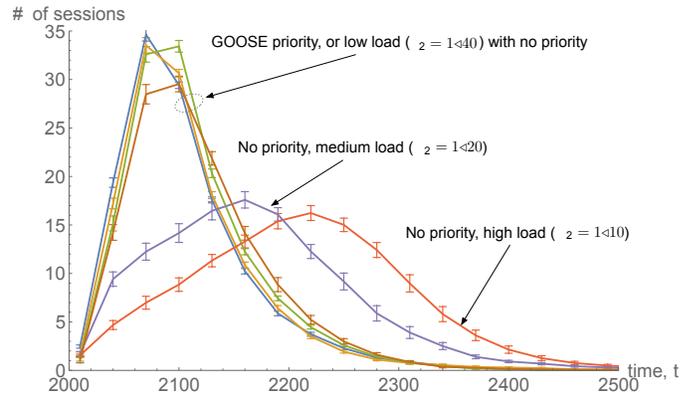}  
    \caption{\small{GOOSE burst periods $\mathcal{T}$ for $\lambda_2=\{1/10,1/20,1/40\}$} [s$^{-1}$].} \vspace{-3mm}
    \label{fig:burst}
\end{figure}

In Fig.~\ref{fig:burst}, we show 
the variation of the GOOSE burst duration ($\mathcal{T}$)  when $\lambda_{2}$ varies, with and without GOOSE traffic priority.
As can be observed, the mean GOOSE burst duration versus the number of GOOSE session curves almost overlap when GOOSE traffic priority is endorsed. This is because GOOSE sessions are guaranteed to obtain sufficient resources and will not be affected by parallel video sessions. Not surprisingly, it is observed that if no priority is given, the mean GOOSE burst duration increases significantly as the video traffic load increases ($\lambda_{2}$ increases).

\subsubsection{Reject, discard, and downgrade ratios}
\begin{figure}[t]
  \centering
  \includegraphics[width=1\columnwidth]{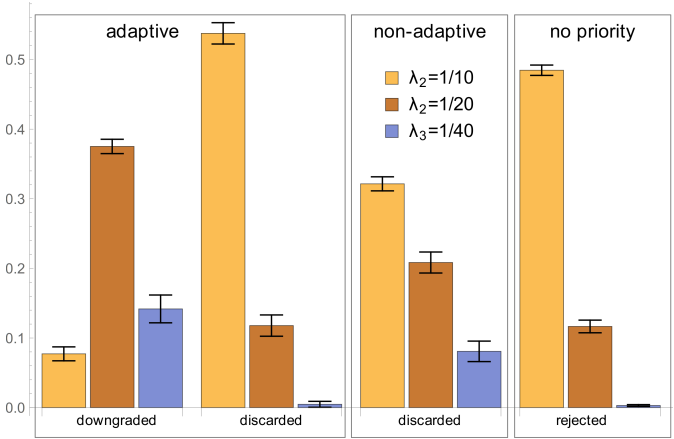}  
  \caption{\small{Reject ($r_{\tt rj}$), discard  ($r_{\tt dc}$), and downgrade ($r_{\tt dw}$) ratios.}} 
  \label{fig:ratios}
\end{figure}

In Fig.~\ref{fig:ratios}, the 
three defined ratios in Subsec.~\ref{subsec:perf} are illustrated. For NC3 (with both priority and adaptive video rate), we observe two ratios for downgraded and discarded video flows, as video sessions will be downgraded first and then discarded if there is still no sufficient resources after downgrading, in order to provide resources to GOOSE sessions. When video traffic load is high in the network, many video sessions would be discarded even before the GOOSE sessions are injected into the system. Thus for $\lambda_{2}=1/10$ [s$^{-1}$], the discard ratio is substantially high whereas the downgrade ratio is comparatively low. 

For NC2 (with priority but non-adaptive video) and NC1 (neither priority nor adaptive),  video sessions would be discarded directly without downgrading when insufficient system capacity occurs. As traffic load increases, more video sessions will be discarded or rejected.

\section{Conclusions and Future Work}\label{sec:con}

In this paper, we have proposed three multi-dimensional Markov models to address the resource allocation problem in 5G RAN slicing using mixed numerologies for smart grid control and protection traffic. Based on the developed models, we  evaluate the transient performance of two traffic types (using different numerology) with and without 
traffic priority and investigate resource utilization with respect to system capacity versus the number of sessions served. Through analysis and simulations, we reveal the behavior of a sliced network with heterogeneous traffic during the transient period and demonstrate the importance of traffic priority as well as the necessity of service adaptation. In addition, we proposed a criterion for selecting the most appropriate numerology in a 5G RAN based on the service requirements. Our study provides a reference framework for 5G cellular operators to increase RAN resource utilization by  flexible allocation of radio resources considering both normal and busty traffic conditions. 

As future work, one may investigate dynamic resource allocation at a frame or subframe level considering hybrid numerologies in a subframe. Initial access considering smart grid traffic priority could be another topic of interest.

\section*{Acknowledgments}
\vspace{-1mm}

This paper has been funded by CINELDI - Centre for intelligent electricity distribution, an 8 year Research Centre under the FME-scheme (Centre for Environment-friendly Energy Research, 257626/E20). The authors gratefully acknowledge the financial support from the Research Council of Norway and the CINELDI partners. For the work of V. Casares-Giner and F. Y. Li, the research leading to these results has received funding from the NO Grants 2014–2021, under Project contract no. 42/2021 - “A Massive MIMO Enabled IoT Platform with Networking Slicing for Beyond 5G IoV/V2X and Maritime Services (SOLID-B5G)." The work of V. Casares-Giner  was supported by Grant PGC2018-094151-B-I00 funded by MCIN/AEI/10.13039/501100011033 and ERDF A way of making Europe.


\end{document}